\documentclass[10pt,journal]{IEEEtran}

\usepackage{graphicx,changepage}
\usepackage{float}
\usepackage{subfigure}
\usepackage{scalerel}
\usepackage{amsmath}
\usepackage{xcolor}
\usepackage{cite}
\usepackage{array}
\usepackage{pifont}
\usepackage{fixltx2e}
\usepackage{relsize}
\usepackage{amssymb}

\usepackage{tikz}
\newcommand*\circled[1]{\tikz[baseline=(char.base)]{\node[shape=circle,draw,inner sep=1pt] (char) {#1};}}

\ifCLASSINFOpdf
\else
\fi

\begin{document}
\title{RF Low Noise Amplifiers and Power Amplifiers using Tunnelling Barrier modulated Superconductor-Semiconductor-Superconductor Junctions}
\author{\IEEEauthorblockN{Debopam Banerjee}\\
\IEEEauthorblockA{Analog Devices, India\\}}
%\textit{debopam.banerjee@analog.com}}}

\markboth{IEEE Journal Name,~Vol.~**, No.~**, Feb~2021}%
{Shell \MakeLowercase{\textit{et al.}}: Bare Advanced Demo of IEEEtran.cls for IEEE Computer Society Journals}


\maketitle

\begin{abstract}
In high-performance RF transceivers the thrust of has been on lower noise LNAs and high-power high-speed PAs. The performance trade-off has forced many solutions to have the LNA and PA on a separate die compared to the rest of baseband. Often fabricated in GaAs or InP, they need to be properly interfaced with the rest of signal chain leading to signal integrity issues. Applications for such technologies range from defence, aerospace to scientific instrumentation. In this paper we propose a device using superconductor-semiconductor-superconductor (SC-Sm-SC) junction with a controllable gate terminal to modulate the tunnelling resistance. This arrangement in a typical resonant-tank LNA or an impedance matching PA would reduce the parasitic capacitances leading to higher frequency operation. Also by virtue of Cooper-pair bosons being the bulk carriers, noise due to carrier-carrier and carrier-lattice scattering will be much lower than its conventional CMOS counterparts. Our calculation shows that it can be used in a LNA with 20dB gain till 36GHz and in a PA to deliver -10dBm power till 350GHz.
\end{abstract}


\IEEEpeerreviewmaketitle

\setlength{\parskip}{1em}
\makeatletter
\newcommand*{\rom}[1]{\expandafter\@slowromancap\romannumeral #1@}
\makeatother

\section{Introduction}
Josephson junctions employing the phenomenon of tunnelling of Cooper-pairs between superconductors through a thin insulator have been around for a long time. While digital switching circuits employing them have been presented as an alternative to CMOS, they have not really found a foothold in the analog domain due to lack of a mechanism to extract any meaningful gain. Similarly junctions/devices employing 2-dimensional electron gas (2-DEG) have been explored in academia but those too need large load resistances to extract any meaningful analog gain. This has a negative effect on the maximum operating frequency of the device and also on the output noise power of the amplifying stage. \par

In conventional silicon based MOSFET LNAs and PAs, the main problems are linearity, power consumption for higher operating frequency and intrinsic noise (thermal and flicker). We will briefly compare them below -:
\begin{itemize}
  \item MOSFETs being square-law devices have an intrinsic $2^{nd}$-order distortion component. Most of the times there is also a $3^{rd}$-order distortion component. The proposed device is just a variable resistor which depends linearly on the controlling voltage. Thus the intermodulation products at LNA output in presence of a jammer would be reduced to a large extent. In the case of PA this means lesser spurious power spilling over to the sidebands requiring lesser filtering.
  \item Square root dependence of $g_m$ on $I_D$ in MOSFETs given by $\sqrt{2\mu_nC_{ox}\frac{W}{L}I_D}$. This means that given a load capacitance, we would need to burn 4$\times$ the current in a current biased device in saturation to get a 2$\times$ improvement in operating frequency. While in the proposed device the controlling gate terminal is the one only parameter controlling the effective ON-resistance and thus the frequency response.
  \item Thermal noise would definitely be lower due to the cryogenic operating temperatures. Moreover, as in a regular superconductor, only those scattering events with energy greater than $\Delta_{SC}$ will cause a change in the current flowing. Flicker noise may however still be present as the intermediate layer between the superconductors will be a doped semiconductor.
\end{itemize}

The paper is arranged in four sections. In section-\rom{2} we will discuss the construction of the proposed device and derivation of the tunnelling current density across the potential barrier. In section-\rom{3} we will look at the effective tunnelling resistance and how it can be controlled via the gate terminal. In section-\rom{4} we will look at some probable circuit implementations and also briefly at the frequency response of such circuits compared to their silicon CMOS counterparts. In section-\rom{5} we will derive the noise performance of such devices.

\section{Proposed Device and Junction Current}
\begin{figure}
\centering
\includegraphics[scale=0.83]{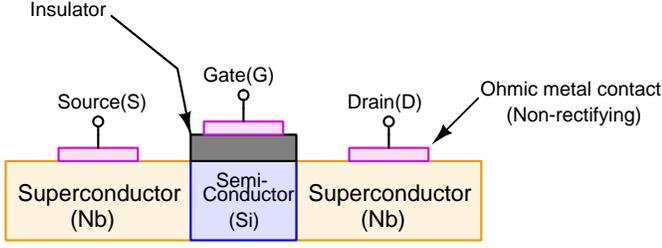}
\caption{Side view of the proposed device showing the thin semiconductor region sandwiched between the superconducting regions.}
\label{device}
\end{figure}

The proposed device is shown in Fig.\ref{device}. It consists of two superconducting layers (can be Niobium) with a thin semiconducting layer (can be silicon) in between. The metallic contacts on the superconducting regions i.e. source and drain connect this device to the rest of the circuit. Thus these contacts need to be ohmic and non-rectifying. The contact on the gate terminal can be made of the same metal as it is non-conducting and is primarily used to generate an electric field in the semiconductor region. The insulator can be $SiO_2$ if the underlying semiconductor is $Si$. Any other insulator which can be easily deposited/manufactured on the thin semiconducting region can also be used.

The device works primarily on the principle of quantum-tunnelling of bosons (Cooper-pairs) through a finite potential barrier which in this case is the intermediate semiconductor layer. Thickness of this layer would typically be $\lesssim$1nm. The gate terminal controls the height of the potential barrier between the source and drain regions. The charge carrier bosons would tunnel from the source to the drain only if there are empty states on the drain side and thus the potential difference between them $V_{DS}$ should be more than $2\Delta_{SC}$. This condition would need to me maintained irrespective of whether the device is current-biased or voltage-biased. The device under such biasing can be approximated as a step voltage extending from the region $-\frac{d}{2}\leq x \leq+\frac{d}{2}$. The depletion region would primarily extend into the semiconductor region as $\vec{E}=0$ inside a superconductor. The depletion region inside the semiconductor can also be reduced to a large extent by increasing the donor concentration. We will now derive the tunnelling current equation through this junction assuming a perfect step potential barrier. \par

The bosonic charge carriers inside a superconductor can be described using a macroscopic wavefunction
\begin{equation}
\Psi(\vec{r},t) = \sqrt{n^*(\vec{r},t)}e^{i\theta(\vec{r},t)}
\end{equation}
where $\Psi$ satisfies a Schr$\ddot{o}$dinger like equation with its modulus giving the density of those carriers $n^*(\vec{r},t)$. The supercurrent equation in an isotropic superconductor with constant $n^*(\vec{r},t)$ is given by
\begin{equation}
\vec{J_S}(\vec{r},t) = -\frac{1}{\Lambda}\bigg[\vec{A}(\vec{r},t) + \frac{\phi_0}{2\pi}\nabla\theta(\vec{r},t) \bigg]
\end{equation}
while the energy of the bosons is related to the phase of the wavefunction as
\begin{equation}
\frac{\partial}{\partial t}\theta(\vec{r},t) = -\frac{1}{\hbar}\bigg[\frac{\Lambda \vec{J_S^2}}{2n^*} + q^*\phi(\vec{r},t) \bigg]
\end{equation}
Assuming $\vec{J_S}$ to be uniform over the junction area and the magnetic vector potential to be zero, we can arrive at the Schr$\ddot{o}$dinger's wave equation for a particle with energy $\mathcal{E}_0$ incident on a potential barrier of magnitude $V_0$ ($V_0 > \mathcal{E}_0$)
\begin{equation}
-\frac{\hbar^2}{2m^*}\nabla^2\Psi(\vec{r}) = (\mathcal{E}_0 - V_0)\Psi(\vec{r}) \qquad for \: |x|\leq a
\end{equation}
where 2a is the width of the semiconducting region. Since the variation is only in the $\vec{x}$-direction, the solution of the above equation reduces to
\begin{equation}
\Psi(x) = C_1cosh(\frac{x}{\zeta}) + C_2sinh(\frac{x}{\zeta})
\end{equation}
Here $\zeta$ is the decay length of the wavefunction inside the semiconductor and is given by
\begin{equation}
\zeta = \sqrt{\frac{\hbar^2}{2m^*(V_0-\mathcal{E}_0)}}
\end{equation}
and thus can be controlled by the gate-to-source voltage $V_{GS}$.

\begin{figure}
\centering
\includegraphics[scale=0.78]{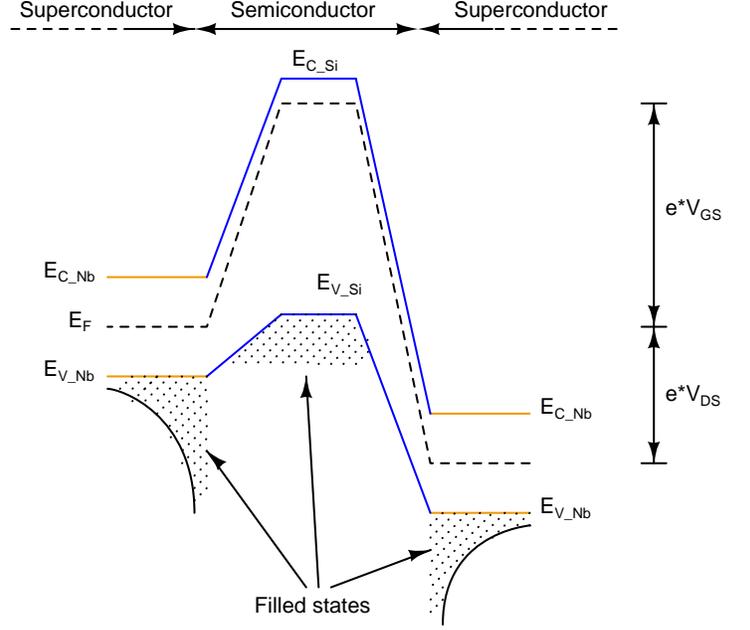}
\caption{Band diagram of the proposed device.}
\label{banddiag}
\end{figure}

We know that a gradient in a wavefunction means an associated particle current density flow exists given by
\begin{equation}
J_S = \frac{q^*}{m^*}Re\bigg(\Psi^*\frac{\hbar}{i}\nabla\Psi\bigg)
\end{equation}
Replacing the candidate $\Psi$ solutions obtained earlier, we get
\begin{equation}
J_S = \frac{q^*\hbar}{m^*\zeta}Im(C^*_1 C_2)
\end{equation}
Working out the boundary conditions gives
\begin{equation}
C_1 = \frac{\sqrt{n^*_1}e^{i\theta_1} + \sqrt{n^*_2}e^{i\theta_2}}{2cosh(\frac{a}{\zeta})}
\end{equation}
\begin{equation}
C_2 = -\frac{\sqrt{n^*_1}e^{i\theta_1} - \sqrt{n^*_2}e^{i\theta_2}}{2sinh(\frac{a}{\zeta})}
\end{equation}
with the final boson charge carriers current density becoming
\begin{equation}
J_S = \frac{e\hbar\sqrt{n_1n_2}}{2m\zeta sinh(\frac{2a}{\zeta})}sin(\theta_1-\theta_2)
\end{equation}
In our working circuits we assume that the device is current biased in such a way that this current density multiplied with the junction area remains higher than the biasing current.

\section{Tunnelling Resistance}
In this section we will briefly review the tunnelling phenomenon and the resulting $I$-$V$ characteristics for temperatures less than critical temperature of the superconductor ($T_C$). As in the previous section, we will assume here too that the device is current biased and the current is very slowly swept from zero to a large value while tracing out the voltage appearing across the device. To begin with the current through the junction being zero, the resultant voltage across the device is also zero. Now when the current is slowly increased to $I_C=J_C \times A_J$ the voltage across the junction still remains zero as the conduction happens via tunnelling of the boson charge carriers (Cooper-pairs). When the current crosses this limit, the normal tunnelling begins and at this very moment the voltage across the junction/device is $2\Delta_{SC}$. This linear part of the curve gives rise to the normal electron tunnelling resistance $R_n$. The origin of these normal fermionic charge carriers can be attributed to the breaking up of the bosonic pairs under the applied field.
\begin{figure}
\centering
\includegraphics[scale=0.7]{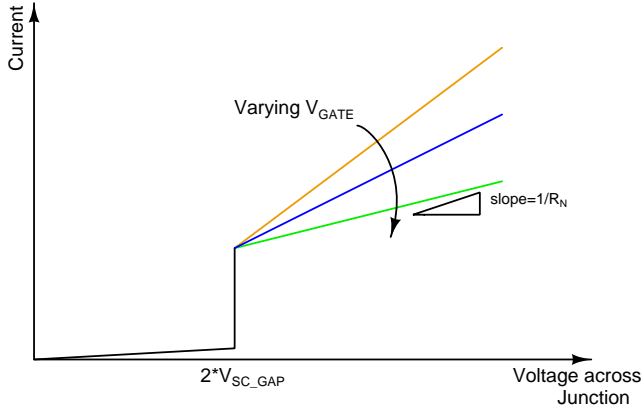}
\caption{Current-Voltage characteristic of the junction inside the proposed device.}
\label{ivchar}
\end{figure}
The parameter $I_C$ is controlled using the $(V_0-\mathcal{E}_0)$ component in $\zeta$. This results in the variable slopes shown in Fig.\ref{ivchar}. If we bias the device using a current source, the load line can be considered as a horizontal straight line at $I_{BIAS}$. With varying $V_{GATE}$ we get a varying $I_C$ which causes the intersection intersection point (biasing point) to shift right or left on the loadline causing a voltage drop across the device. This translates into a gain at the $V_{DS}$ terminals from $V_{GS}$ terminals. \par

The Josephson tunnelling current is given by
\begin{equation}
I = I_Csin(\varphi) \qquad where \: \phi=\theta_1 - \theta_2
\end{equation}
Differentiating this equation with respect to time
\begin{equation}
\frac{di}{dt} = I_C\frac{d\varphi}{dt}cos(\varphi) = \bigg[\frac{2\pi I_C}{\Phi_0}cos\varphi(t)\bigg]v(t)
\end{equation}
The term inside the brackets indicate a time-dependent inductance indicating the component of total tunnelling current consisting of bosons. Typically when this tunnelling current is predominantly due to normal electrons, $\varphi(t)$ can be assumed to be zero with zero mean fluctuations around it. Thus we get
\begin{equation}
\frac{di}{dt} = \frac{2\pi I_C}{\Phi_0}v(t)
\end{equation}
Adding another scattering term to the LHS of above equation to account for the normal electroncs scattering
\begin{equation}
\frac{di}{dt} + \frac{i}{\tau_n} = \frac{2\pi I_C}{\Phi_0}v(t)
\end{equation}
where $\tau_n$ represents the ensemble relaxation time. The resulting conductance $G_n(\omega)$ is given by
\begin{equation}
G_n(\omega) = \frac{2\pi I_C}{\Phi_0}\Bigg(\frac{1}{j\omega + \frac{1}{\tau_n}}\Bigg)
\end{equation}
A thing to note is that the product of this resitance ($R_n$) and $I_C$ is a constant
\begin{equation}
I_CR_n = \frac{\Phi_0}{2\pi \tau_n}
\end{equation}

\section{LNA $\&$ PA circuits}
\begin{figure}
\centering
\includegraphics[scale=0.95]{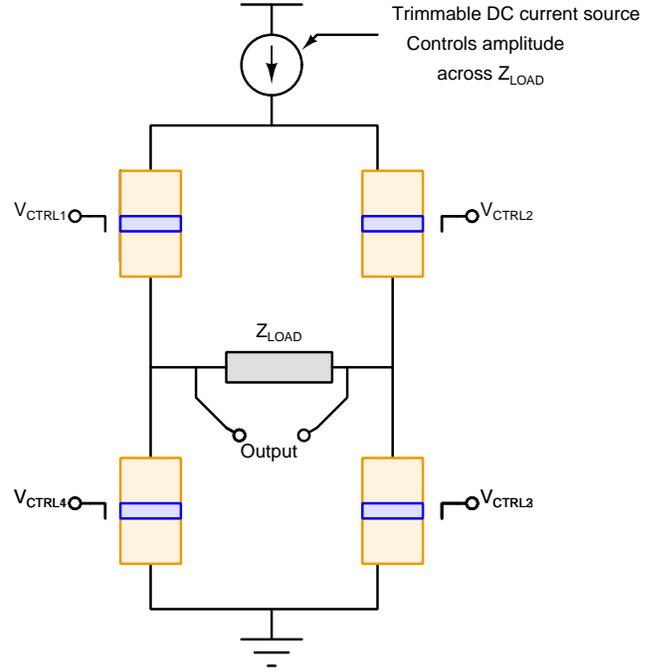}
\caption{The proposed device in a RF PA circuit.}
\label{circuit_pa}
\end{figure}

A probable candidate circuit using the proposed device in a RF-PA is shown in Fig.\ref{circuit_pa}.
Here the circuit simply performs the task of changing the direction of current flow in $Z_{LOAD}$ which might be an interfacing 50$\Omega$ impedance. The switches marked by the controls [1,3] act in unison while the switches marked by controls [2,4] act in unision without any overlap with [1,3]. This resembles a sort of H-bridge structure. \par

Let's assume a typical case of a RF radio transmitter where the maximum output power that we would like to transmit would be around +10dBm. This would correspond to a switched current of 20mA flowing in a 50$\Omega$ load impedance. If we setup the four devices in the four arms such that their $I_C>$20mA then the current would be divided equally into the two arms and none would flow across $Z_{LOAD}$. Next when we want to create a voltage drop of +1V, we would need to lower the $I_C$ of the top-left and bottom-right devices using their gate terminals. This would force the bias current to flow through the top-right device, the load resistance and finally the bottom-left device. For getting a voltage drop of -1V we would need to reverse the ON and OFF devices. Now the top-left and bottom-right devices would be programmed to have higher $I_C$ while the top-right and bottom-left devices would be tuned to have lower $I_C$. \par

Present research in superconducting materials has shown materials that can sustain a $J_C$ of upto $2\times 10^6 A/cm^2$, the area needed to support 20mA current would be as low as $1\mu m^2$. If we assume that the primary capacitance would be from the junction capacitance ($C_{jn}$) between the superconducting electrodes and the semiconductor as the dielectric, we would not be far from actual numbers as there is practically no depletion region to give rise to the depletion sidewall capacitances typically seen in MOSFET devices. We also do not have any oxide overlap $C_{GS}$ or $C_{GD}$ as seen in MOSFETs because $\vec{E}=0$ inside a superconductor. Thus the junction capacitance would be
\begin{equation}
C_{jn} = \frac{\epsilon_r\epsilon_0A_{jn}}{d}
\end{equation}
where $d$ is the width of the semiconductor region. Plugging in the values we get a $C_{jn}=103.4fF$. The time constant of the circuit would be
\begin{equation}
\tau_{RC} = Z_{LOAD} \times \frac{C_{jn}}{2}
\end{equation}
with the factor of 2 coming due to two devices being in series with the load impedance. The effective frequency is given by
\begin{equation}
f_{3dB} = \frac{1}{2\pi Z_{LOAD}\frac{C_{jn}}{2}} = 61.56GHz
\end{equation}
The direct tradeoff between the maximum output power capability with frequency of operation should be clearly discernible from the above example. Another benefit that should be well evident from this implementation is that if the switches would have been implemented using classical MOSFETs with gate control turning ON/OFF to divert the current from one arm to the other, it would have caused some loss in the form of a dc-drop across the ON switches. The magnitude of this loss would be $2R_{ON}\times I_{DC}$. While in the proposed implementation, the voltage appearing across the device is zero due to the tunnelling of bosons (Cooper-pairs). This directly impacts the overall efficiency ($\eta_{eff}$) of the PA.

\begin{figure}
\centering
\includegraphics[scale=0.95]{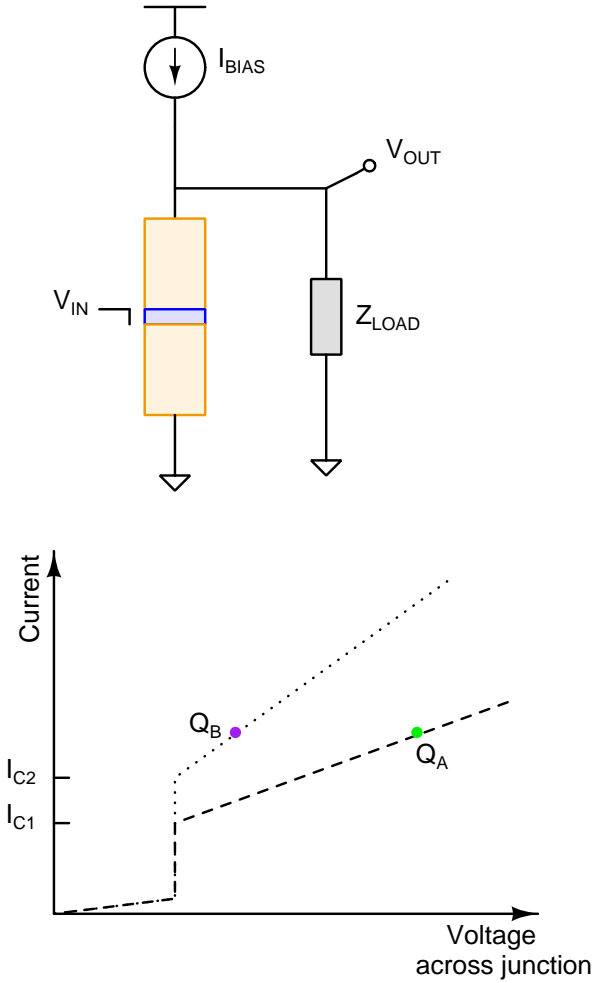}
\caption{The proposed device in a RF LNA circuit.}
\label{circuit_lna}
\end{figure}

A probable candidate circuit using the proposed device in a RF-LNA is shown in Fig.\ref{circuit_lna}. Here a single device is biased by a constant current source which typically might range around 2mA. The $Z_{LOAD}$ here does not need to be a matched 50$\Omega$ as it will primarily be driving a gate-capacitance of a normal MOSFET device as part of a voltage switched mixer or a virtual ground node as part of a TIA+Mixer combined into one. To understand it's working, assume that the input voltage $V_{IN}$ is a two level signal which biases the device in such a way that the critical currents in those cases are $I_{C1}$ when the Q-point is \circled{$\textbf{A}$} and $I_{C2}$ when the Q-point is \circled{$\textbf{B}$}. The transfer-curves in the $I$-$V$ plots in these two cases are shown by dotted and dashed lines respectively. We begin by noting that the output voltages $V_A$ and $V_B$ would correspond to input voltages $V_{AI}$ and $V_{BI}$ and the gain of the circuit is given by
\begin{equation}
A_V = \frac{V_A - V_B}{V_{AI} - V_{BI}}
\end{equation}
The voltages $V_A$ and $V_B$ are given as
\begin{equation}
V_A = 2\Delta + (I_B-I_{C1})R_{n1}
\end{equation}
and
\begin{equation}
V_B = 2\Delta + (I_B-I_{C2})R_{n2}
\end{equation}
The output voltage change is given by
\begin{equation}
V_A - V_B = I_B\times(R_{n1} - R_{n2})
\end{equation}
The term $I_CR_n$ is constant and thus the differential evaluates to zero.
\begin{equation}
V_A - V_B = I_B\frac{\Phi_0m^*}{\pi e^*\tau_n\hbar\sqrt{n_1n_2}A_{jn}} \bigg[\zeta_1 sinh(\frac{2a}{\zeta_1}) - \zeta_2 sinh(\frac{2a}{\zeta_2})\bigg]
\end{equation}
Since for any physically fabricated junction $\frac{2a}{\zeta} >> 1$, we can make the following approximation
\begin{equation}
V_A - V_B = I_B\frac{\Phi_0m^*sinh(\frac{2a}{\zeta_{avg}})}{\pi e^*\tau_n\hbar\sqrt{n_1n_2}A_{jn}} \bigg(\zeta_1 - \zeta_2\bigg)
\end{equation}
where $\zeta_{avg}$ is a rms of $\zeta_1$ and $\zeta_2$. Upon further expansion, we get
\begin{equation}
\begin{split}
V_A - V_B = I_B & \frac{\Phi_0m^*sinh(\frac{2a}{\zeta_{avg}})}{\pi e^*\tau_n\hbar\sqrt{n_1n_2}A_{jn}} \sqrt{\frac{\hbar^2}{2m^*}} \times \\ & \bigg(\frac{1}{\sqrt{V_0-V_{Ai}}} - \frac{1}{\sqrt{V_0-V_{Bi}}}\bigg)
\end{split}
\end{equation}
Denoting
\begin{equation}
\Gamma_{Rn} = \frac{\Phi_0m^*sinh(\frac{2a}{\zeta_{avg}})}{\pi e^*\tau_n\hbar\sqrt{n_1n_2}A_{jn}} \sqrt{\frac{\hbar^2}{2m^*}}
\end{equation}
and doing binomial expansion of the fractions inside brackets, we get
\begin{equation}
V_A - V_B = I_B\Gamma_{Rn}\frac{1}{\sqrt{V_0}} \Bigg[ \bigg(1+\frac{V_{Ai}}{V_0}\bigg) - \bigg(1+\frac{V_{Bi}}{V_0}\bigg) \Bigg]
\end{equation}
Thus the gain of the circuit is finally given as
\begin{equation}
A_V = \frac{V_A - V_B}{V_{AI} - V_{BI}} = \frac{I_B\Gamma_{Rn}}{2V_0^{\frac{3}{2}}}
\end{equation}
This explanation can be easily extended to a bandlimited analog signal with limited swing.

\section{Noise Analysis of the Device}
The main component of noise would be random transitions from the right superconductor to the left superconductor. This would be $f_{BE}(\varepsilon + d\varepsilon)$ which gives the distribution of Bosons (Cooper-pairs), multiplied with $P_R(\varepsilon + d\varepsilon)$ giving the probability of $d\varepsilon$ being occupied on the right superconductor again multiplied by $[1-P_L(\varepsilon + d\varepsilon)]$ giving probability of same energy being vacant on the left superconductor. Also we should multiply by the tunnelling transmission coefficient to get the total number of carriers crossing. Following is the number of bosons available for tunnelling creating a noise current
\begin{equation}
\begin{split}
n^* = \int_{\varepsilon_F+\Delta_{SC}}^{\infty} & \bigg[\frac{2\rho_F\varepsilon sgn|\varepsilon-\Delta_{SC}|}{\sqrt{\varepsilon^2 - \Delta_{SC}^2}}\bigg]    \Bigg[\frac{g_i}{e^{\frac{\varepsilon-\varepsilon_F}{k_BT}}-1}\Bigg] \times  \\ &  \Bigg[\frac{1}{1+\frac{V_0^2}{4\varepsilon(V_0-\varepsilon)}sinh^2[\frac{L}{\hbar}\sqrt{2m^*(V_0-\varepsilon)}]}\Bigg] d\varepsilon
\end{split}
\end{equation}
The first term is the density of states, second term given the probability of occupancy and third term gives the transmission coefficient. Here we have assumed that on an average the bosons would not have a very high energy over ${\varepsilon_F+\Delta_{SC}}$ on the right side and there are a lot of vacant degenerate energy levels just above ${\varepsilon_F+\Delta_{SC}}$ on the left side. Also only $\frac{1}{3}^{rd}$ of these bosons would be directed towards the SC-Sm junction causing the noise current. In the region of integration the term $sgn|\varepsilon-\Delta_{SC}|$ is always positive. $f_c = \frac{k_BT_C}{h}$ gives the maximum soft operating frequency for Niobium as 208.27GHz and with our operating frequency being $\leq f_c$ we get $k_BT=3.313\times 10^{-23}$Joules. Typically the density of charge-carrier bosons $n^*_{SC} = 10^{20}$ in Niobium for $T<T_C$ and $B<B_C$. Below derivation is if we can fit at least $5\times 10^{19}$ in an energy gap of $3.313\times 10^{-23}$Joules above $\varepsilon_F+\Delta_{SC}$.
\begin{equation}
n_1^* = \int_{\varepsilon_F+\Delta_{SC}}^{\varepsilon_F+\Delta_{SC}+\Delta} \frac{2\rho_F\varepsilon sgn|\varepsilon-\Delta_{SC}|}{\sqrt{\varepsilon^2 - \Delta_{SC}^2}} \times g_{\varepsilon_F+\Delta_{SC}} d\varepsilon
\end{equation}
The term $g_{\varepsilon_F+\Delta_{SC}}$ comes from the degeneracy of energy levels available for the bosons to occupy. Using some approximations like $\varepsilon_F >> \Delta_{SC}$ and ignoring smaller terms in binomial expansion, we get
\begin{equation}
n_1^* = 2\rho_F g_{\varepsilon_F+\Delta_{SC}} \frac{\Delta}{\Delta_{SC}} \bigg[1 - \frac{1}{2}\bigg(\frac{\Delta_{SC}}{\varepsilon_F+\Delta_{SC}}\bigg)^2 \bigg]
\end{equation}
Plugging in the values of the constants we can easily get the number of bosons that can be accomodated. Next we will approximate the distribution with the approximation $\varepsilon-\varepsilon_F=\Delta_{SC}+\Delta \approx \Delta_{SC}$ which gives
\begin{equation}
f_{BE}^*(\varepsilon) = \frac{1}{e^{\frac{kT_C}{kT}}-1}
\end{equation}
With $T<T_C$ and a minimum 1GHz operating frequency, we will overestimate the noise component and assume
\begin{equation}
f_{BE}^*(\varepsilon) = \frac{1}{e^{200}-1}
\end{equation}
thus leading to
\begin{equation}
\begin{split}
n^* = \frac{2\rho_F g_{\varepsilon_F+\Delta_{SC}}}{e^{\frac{f_c}{f_{min}}}-1} & \int_{\varepsilon_F+\Delta_{SC}}^{\varepsilon_F+\Delta_{SC}+\Delta} \bigg[\frac{\varepsilon}{\sqrt{\varepsilon^2 - \Delta_{SC}^2}}\bigg]  \\ &  \Bigg[\frac{1}{1+\frac{V_0^2}{4\varepsilon(V_0-\varepsilon)}sinh^2[\frac{L}{\hbar}\sqrt{2m^*(V_0-\varepsilon)}]}\Bigg] d\varepsilon
\end{split}
\end{equation}
The upper limit has been changed to reflect the fact that almost all the tunnelling electrons are concentrated in an energy band of $\Delta$ above $\varepsilon_F+\Delta_{SC}$. In the second term, the argument inside sinh() is a large number for all practical purposes and can be approximated as $sinh^2(\beta)=\frac{1}{4}e^{2\beta}$. Thus the transmission coefficient becomes
\begin{equation}
T_r(\varepsilon) = \frac{1}{1+\frac{V_0^2}{4V_0(\varepsilon-V_0)}\frac{1}{4}e^{2\beta}} , \beta=\frac{L}{\hbar}\sqrt{2m^*(V_0-\varepsilon)}
\end{equation}
\begin{equation}
T_r(\varepsilon) \approx 16 \bigg(\frac{\varepsilon-V_0}{V_0}\bigg)
\end{equation}
Thus finally the expression for carriers becomes
\begin{equation}
n^* = \frac{32\rho_F g_{\varepsilon_F+\Delta_{SC}}}{V_0(e^{\frac{f_c}{f_{min}}}-1)} \int_{\varepsilon_F+\Delta_{SC}}^{\varepsilon_F+\Delta_{SC}+\Delta} \frac{\varepsilon(\varepsilon-V_0)}{\sqrt{\varepsilon^2 - \Delta_{SC}^2}} d\varepsilon
\end{equation}
The definite integral part evaluates to
\begin{equation}
\bigg(\frac{\varepsilon-V_0}{2}\bigg)\sqrt{\varepsilon^2 - \Delta_{SC}^2} - \frac{\Delta_{SC}^2}{3} log(\varepsilon+\sqrt{\varepsilon^2 - \Delta_{SC}^2}) \Bigg\rvert_{\varepsilon_F+\Delta_{SC}}^{\varepsilon_F+\Delta_{SC}+\Delta}
\end{equation}
which can be approximted with the assumptions $\rightarrow$ \\ $eV_0>>\Delta$, $\varepsilon_F+\Delta_{SC}>\Delta$
\begin{equation}
Integral \approx \frac{\Delta}{2}\sqrt{(\varepsilon_F+\Delta_{SC})^2-\Delta_{SC}^{2}} - \frac{\Delta_{SC}^{2}}{3}
\end{equation}
leading to
\begin{equation}
n^* = \frac{32\rho_F g_{\varepsilon_F+\Delta_{SC}}}{V_0(e^{\frac{f_c}{f_{min}}}-1)} \bigg[\frac{\Delta}{2}\sqrt{(\varepsilon_F+\Delta_{SC})^2-\Delta_{SC}^{2}} - \frac{\Delta_{SC}^{2}}{3} \bigg]
\end{equation}
The second part of the noise current, though small, comes from the generation of Cooper-pairs by incident holes/electrons on the semiconductor-superconductor interface resulting in Andreev reflection. The derivation of this noise-current's power spectral density or rms noise is still a work in progress as there is not much literatue available dealing with semiconductor-superconductor junctions under an applied electric field.

\section*{Acknowledgment}
The authors would like to thank... for their valuable inputs.

\bibliographystyle{IEEEtran}

\end{document}